\documentstyle[epsfig]{aipproc}
\input epsf

\begin{document}
\title{First-Principles Studies of Local Order
in Relaxor Ferroelectrics\thanks{To be published in the
AIP proceedings of the Fifth Williamsburg Workshop on First-Principles 
Calculations for Ferroelectrics, February 1998.}}
\author{Mark Wensell and Henry Krakauer}
\address{College of William and Mary\thanks{Supported by the Office of
Naval Research grant N00014-97-1-0049.},
Department of Physics\\P.O. Box 8795,
Williamsburg, VA 23187-8795}
\def\sss{\scriptscriptstyle}

\maketitle

\begin{abstract}
  A key to optimizing the growth of the new single-crystal relaxor
  ferroelectrics\cite{eagle97} is resolving basic questions concerning
  their structural properties and energetics. We report on initial
  first-principles total energy and force calculations, examining the
  energetics of local order in PZN type relaxors.
\end{abstract}

\section*{Introduction}

The optimization of the electromechanical properties of the new
single-crystal relaxor ferroelectrics\cite{eagle97} and the challenge
of growing large crystals raise several fundamental questions. What is
the structure of these materials, and how is it related to their
properties? We first briefly review the present understanding of these
questions and the capabilities of current theoretical approaches in
addressing them.

Although some structural features of relaxor ferroelectrics are
understood, their precise microscopic structure is unknown at the
present time. (This is, of course, even more true of the new single
crystal relaxor ferroelectrics.) The task of determining the atomic
geometry is complicated by the long-range disorder inherent in relaxor
ferroelectrics and the infinite-range Coulomb interaction between
ions, which imposes severe constraints on the structure. It is perhaps
not surprising then that despite years of extensive effort there are
many fundamental questions about the atomic geometry of relaxors that
are still unsettled.

For example, in PMN type relaxors with the stoichiometry Pb(B$^{\sss
  2+}_{\sss 1/3}$B$^{\sss 5+}_{\sss 2/3}$)O$_3$ (here B$^{\sss
  2+}$=Mg, Co, Ni, Zn; B$^{\sss 5+}$=Nb, Ta), the B-sites must
accommodate a 2:1 distribution of cations with very different positive
charges.  Experimental evidence indicates that PMN crystals do this by
forming 1:1 ordered nanometer scale domains dispersed in a disordered
matrix.\cite{cross94,krause79,hilton88,rosenfeld95,viehland93,khachat93}
The ordered domains, probed by diffraction and high-resolution TEM
studies, are seen to have a 1:1 distribution of two different cation
sites (designated $\beta'$ and $\beta''$) arranged on a face-centered
NaCl-type lattice with a doubled perovskite repeat along the [111]
direction. This much seems to be generally accepted. Other details
about the structure are considerably less certain, as discussed below,
with several basic questions that are still not settled: 1) Which
atoms reside on $\beta'$ and $\beta''$ sites?  2) Are the 1:1
nano-domains intrinsically size-limited by the Coulomb interaction or
are they the result of particular crystal growth and annealing
procedures? It is evident that the $\beta'$ and $\beta''$ site
assignments play a crucial role in answering the second question. For
if the nano-domains are not charge-neutral, Coulomb interactions would
be expected to limit the growth of these domains.

One answer to these questions is provided by the widely accepted
space-charge model, which relates the observed relaxor behavior to
postulated nanometer scale concentration
inhomogeneities.\cite{akbas97,cross94,laiho92,sagala92} The
space-charge model postulates that the nano-domains are B$^{\sss 5+}$-
deficient, with the $\beta'$ and $\beta''$ sites occupied exclusively by
the B$^{\sss 2+}$ and B$^{\sss 5+}$ cations respectively, in a NaCl like structure. The
ordered regions thereby carry a net negative charge. This charge
imbalance is then compensated by a B$^{\sss 5+}$-rich disordered matrix. The
resulting large self-Coulomb repulsion of the charged regions would
be expected to limit the size of the nano-domains, and the random
distribution of nano-domain polarizations is an appealing explanation
for the observed diffuse and frequency dependent permitivity of the
relaxor ferroelectrics. Support for this model includes an observed
lack of domain coarsening with long-term annealing in all but the most
recent experiments, discussed below. This has contributed to
widespread acceptance of the space-charge model, despite the absence
of direct evidence for this type of nano-level compositional
segregation.\cite{akbas97,khachat93}

Recently, however, Akbas and Davies have reported experiments on
Pb(Mg$_{\sss 1/3}$Ta$_{\sss 2/3}$)O$_3$ - PbZrO$_3$, in which the size
of fully chemically ordered 1:1 domains (evidenced by x-ray
diffraction) were increased by two orders of magnitude through
annealing conducted at 1325 $^o$C; moreover, fully ordered ceramics
comprised of large domains were found to retain relaxor behavior.
\cite{akbas97} (Previous annealing was restricted to below 970 $^o$C.)
The concentration inhomogeneities of the space-charge model are
inconsistent with these recent results, since Coulomb repulsion would
limit the size of the B$^{\sss 5+}$-deficient 1:1 domains.  There are
several important implications of this recent experiment.  First, the
distribution of B$^{\sss 2+}$ and B$^{\sss 5+}$ concentrations must be
homogeneous at the atomistic scale to permit the growth of large 1:1
domains. Second, the retention of relaxor behavior of the fully
ordered ceramics suggests that the relaxor behavior is due to random
local (at the atomistic scale) polarization fluctuations. Akbas and
Davies proposed a previously dismissed charge-balanced {\it
  random-site} model for cation ordering as a possible candidate for
atomistic scale homogeneity.  According to this scheme, the $\beta''$
sites are occupied exclusively by B$^{\sss 5+}$, while the $\beta'$
sites are occupied by a 50-50 random mixture of B$^{\sss 2+}$ and
B$^{\sss 5+}$ ions.  The structural formula can be represented as
Pb{[B$^{\sss 2+}_{\sss 2/3}$B$^{\sss 5+}_{\sss 1/3}$]$_{\sss 1/2}$
  [B$^{\sss 5+}_{\sss 1/2}]$O$_3$. The random-site model is consistent
  with the domain coarsening observed in this experiment.

It is also consistent with the observations of Teslic {\it et
al}.\cite{teslic97} using pulsed neutron atomic pair-distribution
function measurements of PMN, PZ and PZT, which show that the atomic
environments, particularly that of Pb, are similar in all these
compounds. Teslic {\it et al}. suggest that a large portion of the
ferroelectric polarization is provided by Pb displacements (to
accommodate the lone-pair electrons), and these displacements are
closely coupled to the rotation of the BO$_6$ octahedra. This picture
provides another possible framework for understanding relaxor behavior
at the atomic level. Since the coupling of the Pb polarization to the
environment and the ease of rotation of the BO$_6$ will have different
energy scales, the random distribution of the B cations on the $\beta'$ 
sites might be expected to lead to relaxor behavior.

The above illustrates the status and shortcomings in the present
understanding of the geometry of even the conventional relaxor
ferroelectrics. Even less is known about the single crystal relaxors.
Yet knowledge of the atomic geometry is certainly a necessary
prerequisite for understanding and optimizing their remarkable
properties. A complete description of the atomic geometry will have to
resolve questions like the space-charge versus random-site models. To
do this, local effects such as atomic relaxation (i.e. Pb
displacements and BO$_6$ octahedra rotations) away from ideal
perovskite structure will also have to be treated. Such short-range
order likely plays a crucial role in defining the properties and the
relative stability of different structures. If the atomistically
homogeneous models of the relaxors are in fact correct, then this is
likely to be the case, because in this scenario the nano-domains are
not intrinsically limited in size by Coulomb effects, but inhibited in
growth by geometrical frustration, similar to spin glasses.
\cite{teslic97} Indeed, the role of PbTiO$_3$ admixture in PMN-PT
single-crystal relaxors may be to reduce frustration and facilitate
single-crystal growth.

A powerful theoretical approach to characterize the short-range order
is to use first-principles calculations to determine local atomistic
structures and resolve the energetics of the space-charge versus
random-site model. In this initial study, we employ supercells
containing 15 - 30 atoms to model the short-range order. To
efficiently handle the computational burden incurred in such large
scale calculations, we employ our first-principles mixed-basis
projector method.\cite{singhmb92,singhmb93} 

Looking ahead, we eventually want to model growth processes as well as
long-range disorder and domain interactions. A direct first-principles
approach, however, would require very large simulation cells, a task
that exceeds the capability of any presently available
first-principles method. Nevertheless, complementary methods have been
developed to extend the reach of the first-principles methods.  One
such method is based on using first-principles derived effective
Hamiltonians. These Hamiltonians act in a reduced sub-space of the
full Hilbert space, retaining only the most important degrees of
freedom. 

Molecular dynamics and Monte Carlo simulations with such effective
Hamiltonians have demonstrated their ability to successfully predict
phase transitions and temperature-dependent static and dynamic
properties for the pure constituents systems like BaTiO$_3$, KNbO$_3$,
and PbTiO$_3$.\cite{rabe95,krakauer97} Thus, the correct phase
sequence of ferroelectric phase transitions was obtained in these
materials, showing that the first-principles effective Hamiltonian
used in the simulations captures the essential behavior of the
microscopic fluctuations driving the transitions. In KNbO$_3$ and
BaTiO$_3$ molecular dynamics simulations have revealed the existence
of localized preformed dynamic chain-like structures, which are
present even in the high-temperature paraelectric phase, well above
the cubic-tetragonal phase transition.\cite{krakauer97} The molecular
dynamics simulations also reproduced the essential features of the
diffuse x-ray scattering measurements and the weak temperature
dependence of measured diffuse streak patterns. These studies provided
a framework for understanding both the displacive and order-disorder
characteristics of these phase transitions as both arising from the
softening of an {\it entire branch} of unstable transverse-optic
phonon modes. 

We hope that the present calculations for PZN can lay the foundation
for extending such finite temperature studies to the relaxor
ferroelectrics. Extending the effective Hamiltonian approach to alloy
systems presents many challenges, and may ultimately be unsuccessful.
In that case other approaches, such as effective interatomic
potentials or generalized shell models may be better suited to the
problem. In any case, any of these complimentary methods will be based
on first-principles results of the type presented below.

\section*{Methodology}
Self-consistent local density functional approximation (LDA)
calculations were carried out using a first-principles mixed basis
projector method.\cite{singhmb92} The method uses a Kerker type
pseudopotential to remove the chemically inert innermost core-electron
states from the Hilbert space. For the 30-atom supercells, there were
134 occupied bands, and the basis set consisted of about 3600 plane
waves, corresponding to about a 19 Ry kinetic energy cut-off. This
relatively low kinetic energy cut-off is achieved by the inclusion of
148 local basis functions, including Pb 5$d$, Zn 3$d$, Nb 4$s$, 4$p$,
4$d$, and O 2$s$, 2$p$ orbitals. The semicore Nb 4$s$ and 4$p$ states
were retained in the variational basis, due to their spatially
extended character. The method benefits from Car-Parinello CPU-time
scaling, facilitating the treatment of these large supercells. It
achieves this using fast Fourier transforms and the fact that the
local orbitals are strictly confined within non-overlapping muffin-tin
spheres. Thus there are no two- or three-center integrals, and the
local orbitals overlap only with the plane waves. The muffin-tin
sphere radii were 2.0, 1.9, 1.6, and 1.65 a.u.  for Pb, Zn, Nb, and O
atoms, respectively. In these initial calculations, we chose the
volume to correspond to the experimentally observed volume of
PbTiO$_3$, but allowed the internal atomic positions to relax to the
lowest energy.  Calculations were carried out for 30- and 15-atom
supercells, using one and two k-points, respectively. This corresponds,
roughly, to a 2x2x2 Monkhorst-Pack mesh.

\section*{Results and Discussion}
We performed calculations on five different structural models, based
on [111] B cation-plane stacking. Two are based on [111] planes
containing either Zn or Nb.  The first is a 30 atom model that has the
stacking sequence Zn, Nb, Zn, Nb, Nb, Nb,Zn ... , and we refer to this
model as ``6spcchrg'', containing 6 perovskite units. It features the
1:1 ordering proposed by the space charge model and in addition, crudely
simulates a Nb rich region. The second is a 15 atom model,
``3spcchrg'', with the stacking sequence Zn, Nb, Nb .... We have
imposed {\it R3m} space group symmetry on both of these models.

In the remaining three models the [111] B-site planes do not all
contain pure Zn or Nb. There are two 30 atom models, each of which is
based on the stacking sequence $\beta'$ $\beta''$ ..., where $\beta'$
is [Zn$_{\sss 2/3}$Nb$_{\sss 1/3}$]$_{\sss 1/2}$ and the $\beta''$ site
is pure Nb. These two differ by the in-plane lattice vectors and have
different symmetry. One of these, ``6tria'', has symmetry {\it Imm2} and is based 
on a triangular arrangement of the Zn and Nb atoms. The other, ``6lin'', has
symmetry {\it Fmm2} and features adjacent rows of Zn atoms separating rows of 
Nb atoms. The last structural
model is a 15 atom model patterned on 6tria with each [111] B-site
plane containing [Zn$_{\sss 1/3}$Nb$_{\sss 2/3}$], which we name ``3tria''.

\begin{figure}
\centering
\epsfxsize=4.0 in 
\rotatebox{-90}{ \epsfbox[104 84 543 610]{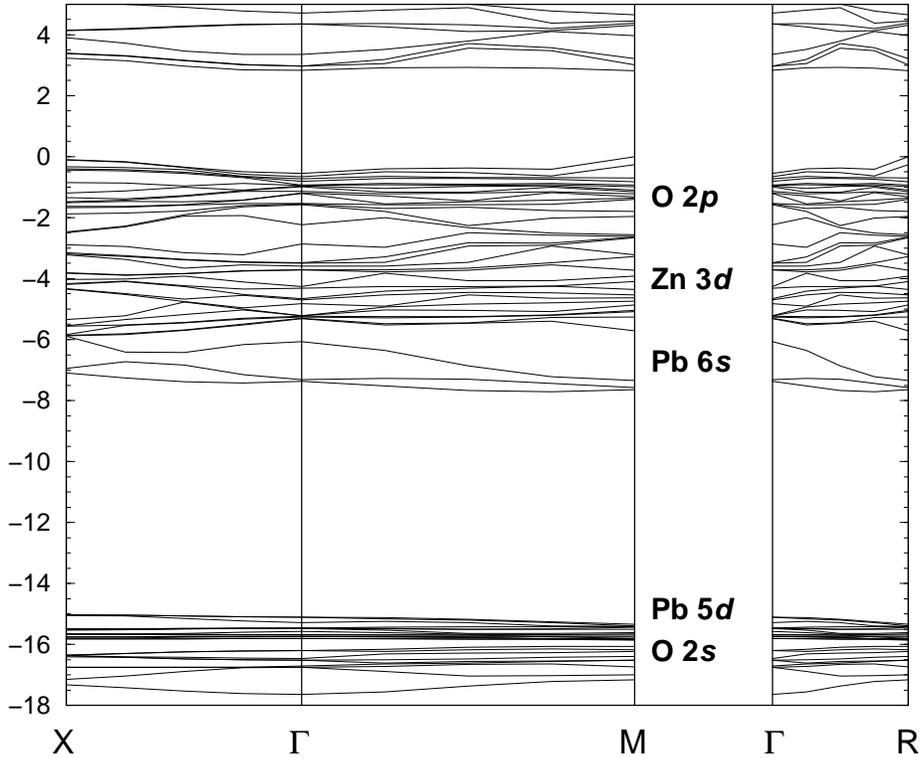} }
\caption{Band structure of 15 atom ``3spcchrg'' cell}
\end{figure}

\begin{figure}
\centering
\epsfxsize=3.4 in
\rotatebox{-90}{ 
\epsfbox[104 100 576 651]{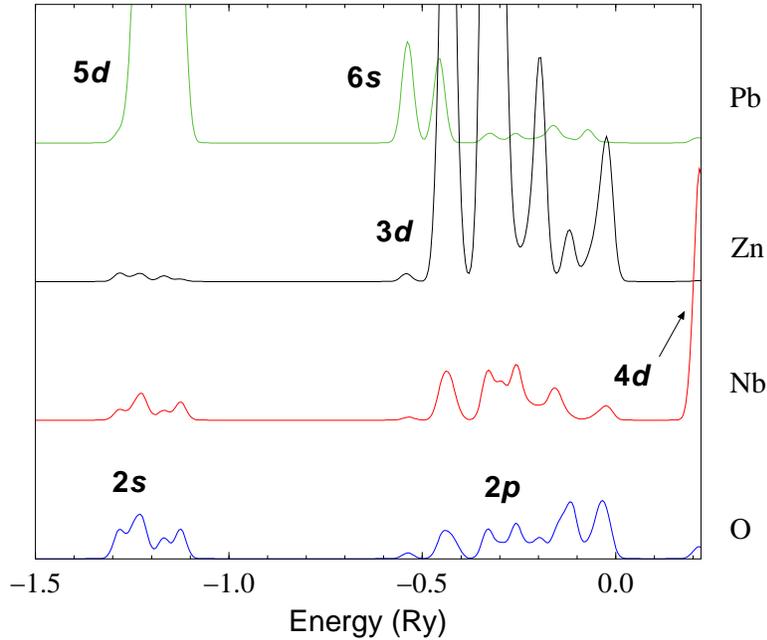}
}
\caption{Density of states of 15 atom 3-space-charge cell.  The
$s-p-d$ character of Pb, Zn, Nb and O were determined using
sphere sizes of 2.0, 1.9, 1.6 and 1.65 bohr, respectively.}
\end{figure}

We begin by presenting the general features of the electronic
structure of these models. Figure 1 shows the band structure for
3spcchrg along high-symmetry directions in the Brillouin zone,
referenced to the ideal PbTiO$_3$ Brillouin zone, and Figure 2 shows
the corresponding density of states projected onto the various atomic
species. Strong Zn 3$d$ - O 2$p$ bonding is evident in these figures.
Other characteristic features of PbTiO$_3$\cite{cohenkrak92}, are also
evident, such as strong Nb 4$d$ - O 2$p$ bonding and Pb 6$s$ - O 2$p$ bonding.

\begin{figure}
\centering
\epsfxsize=4.5 in
\rotatebox{-90}{ 
\epsfbox[46 108 570 671]{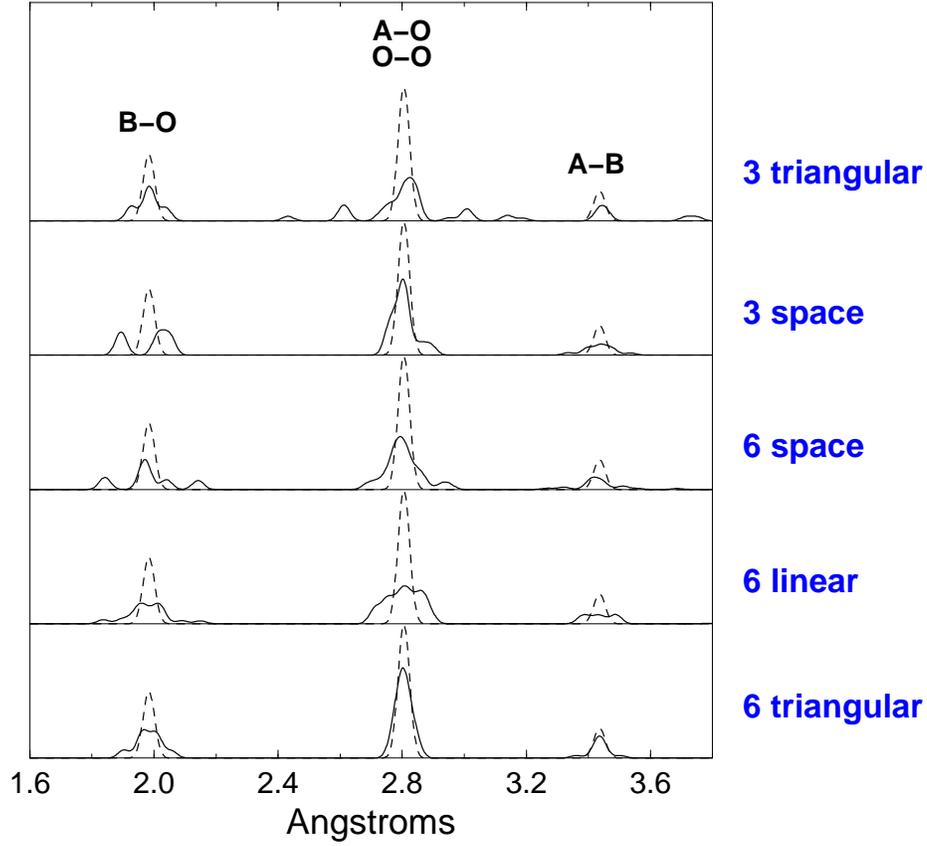}
}
\caption{Pair distribution functions for relaxed and unrelaxed PZN structures}
\end{figure}

\renewcommand{\arraystretch}{1.2}
\begin{figure}

\begin{tabular}{|l|c|c|c|c|c|} 
\hline
\multicolumn{1}{|p{.75 in}|}{\hfil \ \hfill}  
& \multicolumn{3}{c|}{Distances (\AA)}
& \multicolumn{1}{c|}{E$_T$}\\ \cline {2-4}
& \multicolumn{1}{p{1.0 in}|}{\hfil Pb-O \hfill} 
& \multicolumn{1}{p{1.0 in}|}{\hfil Nb-O \hfill} 
& \multicolumn{1}{p{1.0 in}|}{\hfil Zn-O \hfill}   
& \multicolumn{1}{p{.75 in}|}{\hfil (mRy) \hfil} \\
\hline
6-triangle & 2.79--2.83 & 1.90--2.06 & 2.01 & 0  \\
\hline
3-triangle & 2.43--3.19 & 1.93--1.98 & 1.98--2.04 & +30  \\
\hline
3-space & 2.76--2.86 & 1.89--2.05 & 2.02 & +39  \\
\hline
6-space & 2.69--2.95 & 1.84--2.14 & 1.98--2.04 & +63  \\
\hline
6-linear & 2.71--2.87 & 1.84--2.15 & 2.01--2.09 & +73 \\
\hline
Ideal & 2.81 & 1.98 & 1.98 & --  \\
\hline
Expr.$^\dag$ & 2.38, 2.40$^*$ & 1.92--2.12 & 2.03 & --  \\
\hline
\end{tabular}
{\footnotesize {$^\dag$I-Wei Chen, {\it et al.}, 1995}}

{\footnotesize {$^*$Egami {\it et al.}, 1995}}
\begin{center}
{\footnotesize \bf TABLE 1. \rm Summary of PZN bond-lengths and total energies}
\end{center}
\end{figure}

The distribution of nearest neighbor distances is presented in Figure
3 as pair distribution functions, artificially gaussian broadened. The
dashed curves represent the ideal perovskite structure.  There is
large differences in the distribution of nearest neighbor distances in
the different structures. The ranges of values are also tabulated in
Table I.  The characteristic ferroelectric splitting of Nb-O
bondlengths is evident in all the structural models, with one bond
shorter than ideal and the other longer. The widest variation occurs
in the 6lin model. The Zn-O distances are all closely clustered around
the ideal value, the widest variation again occuring for the 6lin
structure. By contrast with Nb-O, however, the Zn-O distances are
mostly equal to or larger than the ideal perovskite B-O distance.  By
far the largest differences from ideal occur for the Pb-O bond
lengths. The widest variation occurs for the 3tria structure. In all
cases, the lower end of the range is less than the ideal value, while
the upper end is larger than ideal. Interestingly, the smallest
variation occurs for the lowest energy 6tria structure. The ranges of
values for the B-O distances agree well with the XAFS measurements of
Chen {\it et al.}\cite{chen96} for PZN, but less well for the Pb-O
bond length (only the shortest measured Pb-O bond was reported).  The
exception is the 3-tria model, whose smallest value is 2.43 \AA.

The total energies (per 15 atom PZN formula unit) of each of the five
structures are compared in Table I, with all energies referenced to
the 6tria model, which had the lowest total energy. The next most
stable structures are the 3tria and 3spc models, with the 6lin and
6spc being least stable. This suggests that the random-site model (in
the 6tria form) has the lowest energy. However, we must still investigate the
effect of increasing the k-point sampling and reducing the imposed
symmetry. 

\section*{Conclusions}

We have performed first-principles total energy and force studies of
models of the solid solution relaxor ferroelectric PZN. Five different
models were investigated incorporating the effects of [111] B-site
ordering in both the space charge and random-site models. In all the
calculations sizeable Z-O bonding occurs, but the Z-O bond distances
are all close to the ideal perovskite value. Comparisons of fully
relaxed total energies in these initial calculations suggest that the
random site model has the lowest energy. The effects of denser
k-point sampling and reducing the imposed symmetry have yet to be
investigated, however. Nb-O and Zn-O distances are consistent with XAFS
measurements but the measured Pb-O bond length is significantly
smaller than all but one of the models.  Detailed comparisons with
neutron pair-distribution measurements would be desirable.

\end{document}